\def\year{2020}\relax
\documentclass[letterpaper]{article}

\usepackage{aaai20}
\usepackage{times}
\usepackage{epsfig}
\usepackage{graphicx}
\usepackage{amsmath}
\usepackage{amssymb}
\usepackage{bm}
\usepackage{multirow}
\usepackage{array}
\usepackage{booktabs} 
\usepackage{url}  
\usepackage{comment}
\newcolumntype{M}[1]{>{\centering\arraybackslash}m{#1}}
\newcolumntype{P}[1]{>{\centering\arraybackslash}p{#1}}

\title{Precipitation Forecasting via Multi-Scale Deconstructed ConvLSTM}

\author{Xinyu~Xiao\textsuperscript{1}, Qiuming Kuang\textsuperscript{2}, Shiming~Xiang\textsuperscript{1}, Junnan Hu\textsuperscript{2}, Chunhong~Pan\textsuperscript{1}\\
	\textsuperscript{1}{National Laboratory of Pattern Recognition, Institute of Automation, Chinese Academy of Sciences, Beijing, China}\\
	\textsuperscript{2}{Meteorological Service Center, China Meteorological Administration}\\
	\{xinyu.xiao, smxiang, chpan\}@nlpr.ia.ac.cn, kuangqiuming2008@163.com, hujunnan2@126.com}

\begin{document}
	

	\maketitle

\begin{abstract}
	Numerical Weather Prediction (NWP), is widely used in precipitation forecasting, based on complex equations of atmospheric motion requires supercomputers to infer the state of the atmosphere. Due to the complexity of the task and the huge computation, this methodology has the problems of inefficiency and non-economic. With the rapid development of meteorological technology, the collection of plentiful numerical meteorological data offers opportunities to develop data-driven models for NMP task. In this paper, we consider to combine NWP with deep learning. Firstly, to improve the spatiotemporal modeling of meteorological elements, a deconstruction mechanism and the multi-scale filters are composed to propose a multi-scale deconstructed ConvLSTM (MSD-ConvLSTM). The MSD-ConvLSTM captures and fuses the contextual information by multi-scale filters with low parameter consumption. Furthermore, an encoder-decoder is constructed to encode the features of multiple meteorological elements by deep CNN and decode the spatiotemporal information from different elements by the MSD-ConvLSTM. Our method demonstrates the data-driven way is significance for the weather prediction, which can be confirmed from the experimental results of precipitation forecasting on the European Centre Weather Forecasts (EC) and China Meteorological Forecasts (CM) datasets.
	
\end{abstract}

\section{Introduction}
Precipitation forecasting \cite{HernandezSJCD16,sloughter2007} is an important component in the normal functioning of human society, which is very significance for transportation, agriculture, manufacturing and public safety, etc \cite{ZhengCWY14a}. Therefore, it has large requirements for real-time and accuracy. However, the atmospheric movement is very complex, the precipitation conditions are inferenced by the varying of the various meteorological elements, the regularity of precipitation is difficult to master. The precipitation forecasting is still a great challenge to scientists.

Currently widely used method of precipitation forecasting is numerical weather prediction (NWP) \cite{navon2009data} which analyzes the corresponding meteorological elements like divergence, fraction of cloud cover, geopotential, ozone mass mixing ratio, potential vorticity, relative humidity, specific, cloud ice water content, specific cloud liquid water content, specific humidity, temperature, horizontal(U) component of wind, vertical(V) component of wind, vertical velocity and vorticity (relative), etc. to capture the state of the atmosphere. This kind of method is established based on the flow mechanics and thermodynamics principles which describe the laws of atmospheric motion needs massive data to collect and supercomputer to calculate. This expensive computing takes up a lot of social resources and affects the further development of precipitation forecasting.

With the rise of the wave of deep learning, some researchers attempt to utilize the deep neural network for precipitation forecasting \cite{HernandezSJCD16}. The motivation is to learn the regular pattern of precipitation in a data-driven way which avoids solving the complex equations of atmospheric motion. Shi et al. \cite{ShiCWYWW15} introduced the radar echo maps as the sources and proposed a convolutional LSTM (ConvLSTM) based model for precipitation nowcasting. However, there are no causal physical connections between the radar echo maps and the precipitation. How to combine NWP with the advanced machine learning methods to achieve a more efficient and effective solution is an urgent problem to be solved.

The data of meteorological elements has spatiotemporal property which can be processed by ConvLSTM. However, there are several major limitations of the current structure of ConvLSTM. Firstly, the huge parameter size of ConvLSTM limits the computational efficiency of it. The operation of all the gates in ConvLSTM adopts the same convolutional structure which will bring in a lot of computational consumption. Secondly, the convolutional structure in ConvLSTM limits the feature fusion of the spatiotemporal data. Single kernel size of the filter in the convolutional structure makes it impossible for the ConvLSTM to extract rich contextual information to fusion multiple features.

To efficiently and effectively combine NWP with deep learning, this paper propose an encoder-decoder architecture with multi-scale deconstructed ConvLSTM (MSD-ConvLSTM) for precipitation forecasting. At first, to reduce the parameters and improve the fusion capacity of ConvLSTM, deconstructing the original convolutional structure to a fully-connected (FC) layer and a single channel convolution to build a new structure of the gate. Then, depending on the structure of the deconstructed ConvLSTM, the MSD-ConvLSTM which applies a new CNN module (mConv) with multi-scale filters is proposed. The mConv is applied on the input modulation gate to capture the contextual feature information at multiple scales. An encoder-decoder architecture of deep CNN+MSD-ConvLSTM is proposed. In the architecture, a sequence of the numerical values of multiple meteorological elements are fused in an effective way to forecast the precipitation in over short period of time. The experiments for the future rainfall forecasting on the EC and CM datasets demonstrate the effectiveness of our method.

The main contributions of this paper are:
\begin{itemize}
	\item[-] Depending on the evaluation of the deconstruction of the original convolutional structure to a FC layer and a single channel convolution, the introduction of the multi-scale filters in convolution, a multi-scale deconstructed ConvLSTM (MSD-ConvLSTM) is proposed. 
	\item[-] An encoder-decoder architecture combines NWP with deep neural network is designed. The deep CNN encoder is to encode the numerical values of multiple meteorological elements and MSD-ConvLSTM decoder is to fuse the contextual information of the extracted features. The architecture provides an efficient and effective data-driven way in precipitation forecasting.
	\item[-] This model proposes a valuable framework to handle precipitation forecasting problem. The excellent experimental results of our model demonstrate that artificial intelligence can provide efficient and effective solutions for weather forecasting.
\end{itemize}

\section{Related Works}

Here, we first review the relevant studies of precipitation forecasting, then review the researches of recurrent neural networks for spatiotemporal modeling.

\vspace{-0.1cm}
\subsection{Relevant Studies of Precipitation Forecasting}

Including precipitation forecasting, the studying of weather forecasting has been developed several centuries \cite{hall1999,gope2016early,gneiting2005weather,campbell2005weather,MaqsoodKA04}. The mainstream approaches \cite{richardson2007weather,tolstykh2005some} are NWP-based which simulate atmospheric motion by numerical analyzing. These methods consume expensive computing. Some researchers attempted to learn the regularity of weather system by traditional machine learning methods like SVM \cite{SapankevychS09}, ARIMA \cite{chen2011comparison} and so on. But the modeling capability limits the progression of these methods. The developing of deep learning allows the reliability of this data-driven way showing prospect to solve weather forecasting problem. In precipitation forecasting, Emilcy et al. \cite{HernandezSJCD16} introduced a precipitation forecasting model which includes an autoencoder for reducing and capturing non-linear relationships between attributes and a multilayer perceptron to predict the accumulated daily precipitation for the next day. Shi et al. \cite{ShiCWYWW15} taken the precipitation forecasting as a spatiotemporal sequence forecasting problem and applied the ConvLSTM as the translator. The method uses radar echo sequences for model training.

\vspace{-0.1cm}
\subsection{RNNs for Spatiotemporal Modeling}

Recurrent neural networks \cite{GravesMH13} have been widely used to process sequential data. To modeling the spatiotemporal data, several variants of LSTM have been proposed \cite{ShiCWYWW15,AlahiGRRLS16}. The convolutional LSTM (ConvLSTM)) \cite{ShiCWYWW15} replaces the full-connection with convolution is proposed to process sequential radar echo maps for precipitation forecasting. To process the scenes with spatiotemporal structure, structural-RNN \cite{JainZSS16} combines the power of high-level spatiotemporal graphs and sequence learning success of RNNs. Zhang et al. \cite{0010ZMSSB18} proposed several variants of ConvLSTM to explore the effects of convolutional structures and attention mechanism in ConvLSTM. From the results of experiments, they found that neither the convolutional structures in the three gates of ConvLSTM nor the extra spatial attention mechanisms contribute in the performance improvements.

In this paper, we explore the deconstruction and multi-scale of convolution to propose a MSD-ConvLSTM. Then, the MSD-ConvLSTM is applied to design a forecasting model of precipitation, which can efficiently and effectively process the spatiotemporal meteorological elements to introduce the deep learning for numerical weather predicting.

\begin{figure*}[t]
	\begin{center}
		\includegraphics[scale=.75]{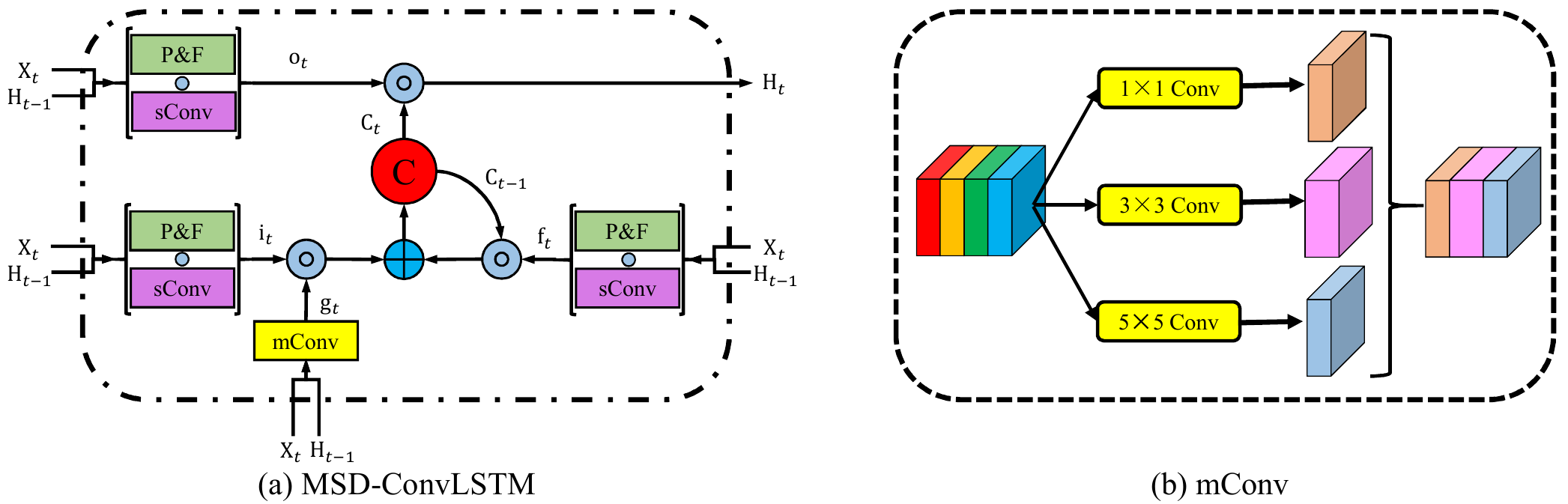}
		
	\end{center}
    \vspace{-0.6cm}
	\caption{The architecture of the multi-scale deconstructed ConvLSTM. (a) denotes the whole MSD-ConvLSTM. (b) is an example of the CNN module with multi-scale filters. ``P\&F" is combining the spatial global average pooling with the fully-connected layer. ``sConv" denotes the single channel convolution. ``mConv" refers to the proposed CNN module with multi-scale filters in equation (\ref{g}). Here, the basic kernel size $K$ is set to 3. ``$\circ$" means the hadamard product. }
	\label{fig:lstm}
	\vspace{-0.6cm}
\end{figure*}

\section{Method}
In this section, we first propose the deconstruction of ConvLSTM to improve its computing efficiency for numerical weather data. Then, multi-scale filters are introduced to propose a Multi-Scale Deconstructed ConvLSTM (MSD-ConvLSTM) which can increase the receptive field area of weather feature. The details of an encoder-decoder architecture combines NWP with deep neural network are described. Finally, the learning method of this model is introduced.

\subsection{Multi-Scale Deconstructed ConvLSTM}

\subsubsection{Deconstruction of ConvLSTM Convolution}

Through the ConvLSTM equation \cite{ShiCWYWW15}, the parameter size of ConvLSTM is large which can be calculated as follows:
\begin{align}
\label{pa}
\Theta_{ConvLSTM} = K\times K\times(C_x+C_h)\times C_h \times 4,
\end{align}
where $K$ is the size of the convolutional kernel in ConvLSTM units; $C_x$, $C_h \gg 1$ are the channel length of the input $\mathbf{X}_{t}$ and hidden state $\mathbf{H}_t$. It indicates that the parameter size of ConvLSTM occupies a lot of memory space and limits the computational efficiency of the model.

The functions of the three gates (input gate, output gate and forget gate) in ConvLSTM are to selectively measure the current input or output, selectively forget the previous memory, respectively. According to Zhang et al. \cite{0010ZMSSB18}, the effect of the convolutional structures of the three gates is not always remarkable. They apply the selective measurement on the channel-wise calculation and replace the convolutional structures of the gates by the FC layers. However, for many tasks such as object detection \cite{DaiLHS16}, semantic segmentation \cite{0005DSZWTA18} and weather prediction \cite{ShiCWYWW15}, etc., the capture of the contextual information between the pixel-level features is the critical factor in learning. Depending on the functions of the gates, we find that the traditional convolutional structure in the ConvLSTM is to measure the feature information on the spatial and channel-wise level, respectively. Therefore, we propose a deconstructing mechanism of the original convolutional structure, which can be seen in Fig. \ref{fig:lstm} (a). 

The deconstruction is to decompose the convolutional structure to channel-wise and spatial operation modules. Specifically, each convolution in the three gates is decomposed to a ``P\&F" module and a single channel convolution (sConv). The ``P\&F" module combined by a spatial global average pooling with a fully-connected layer is to calculate the channel-wise feature information. Taking the input gate as example, the module can be formulated as follows:
\begin{eqnarray}
\label{PF}
\bar{\bm{x}}_{t}\hspace{-0.2cm}&=&\hspace{-0.2cm}GAP(\mathbf{X}_{t}),\nonumber\\
\bar{\bm{h}}_{t-1}\hspace{-0.2cm}&=&\hspace{-0.2cm}GAP(\mathbf{H}_{t-1}),\nonumber\\
\bm{i}_{t}^{c}\hspace{-0.2cm}&=&\hspace{-0.2cm}\sigma(\mathbf{W}_{xi}\bar{\bm{x}}_{t} + \mathbf{W}_{hi}\bar{\bm{h}}_{t-1} + \mathbf{b}_{i}),
\end{eqnarray}
where $GAP$ denotes a global average pooling layer.  $\bm{i}_t^{c} \in \mathbb{R}^{C_h}$ is vector and computed by a FC layer. Meanwhile, the sConv module is a dimensional reduction convolution in which the channel size of the output is reducing to 1. The learned spatial weights of each gate are uniformly applied to each channel of the input and hidden state. The updating procedure in input gate is computed below:
\begin{eqnarray}
\label{single}
\mathbf{i}_{t}^{s}\hspace{-0.2cm}&=&\hspace{-0.2cm}\sigma(\mathbf{W}_{xi}\ast\mathbf{X}_{t} + \mathbf{W}_{hi}\ast\mathbf{H}_{t-1} + \mathbf{b}_{i}),
\end{eqnarray}
where $\mathbf{i}_{t}^{s} \in \mathbb{R}^{1\times W\times H}$. Comparing with original convolution, the deconstruction of the gates into spatial and channel-wise levels reduces great consumption of parameters.

\subsubsection{Multi-Scale Deconstructed ConvLSTM}

\begin{figure*}[t]
	\begin{center}
		\includegraphics[scale=.8]{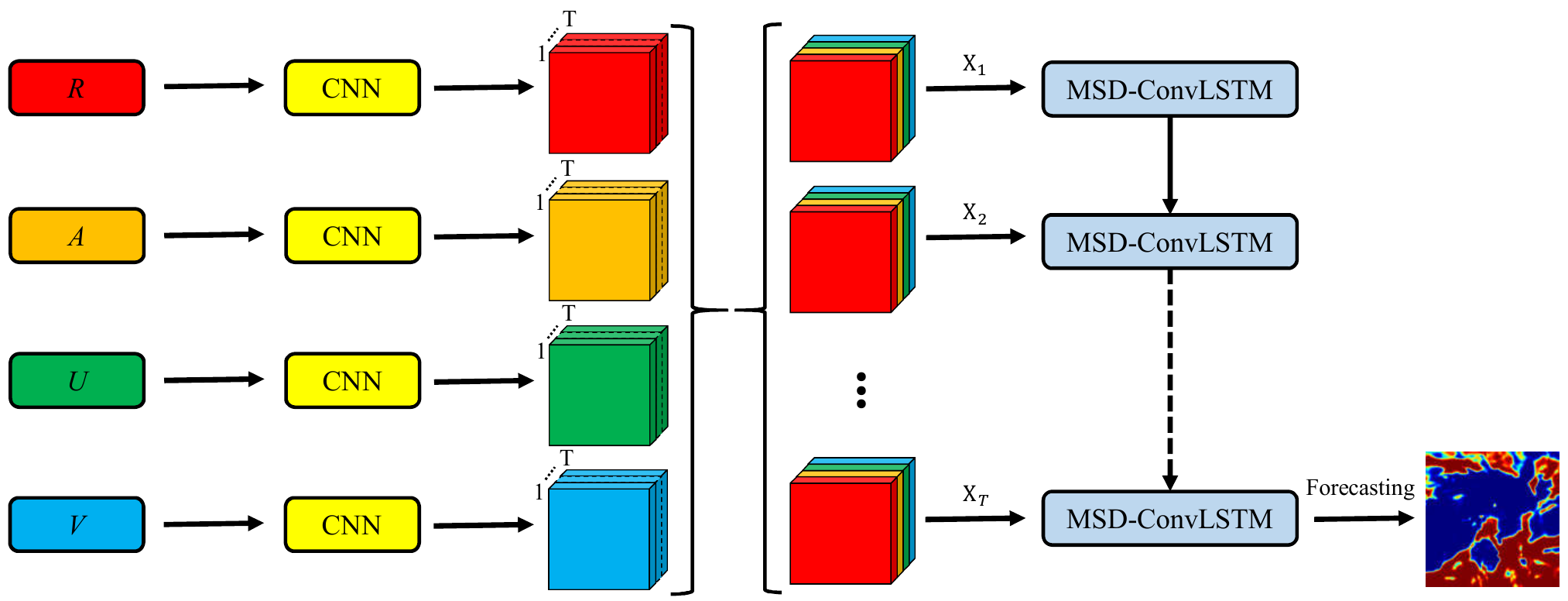}
		
	\end{center}
	\caption{The architecture of our proposed encoder-decoder. The CNN encoder extracted features of different meteorological elements are concatenated at the corresponding time step. The proposed MSD-ConvLSTM is used as the decoder. ``R'' is relative humidity, ``A'' denotes air temperature, ``U'' means U component of wind and ``V'' is V component of wind.}
	\label{fig:model}
	\vspace{-0.5cm}
\end{figure*}

According to \cite{ChenZPSA18}, the capture of the contextual information of each element determines the performance of the pixel-wise prediction. Adjusting filter¡¯s field-of-view to calculate the multi-scale information of the affinity field can capture the semantic relations between neighboring pixels. Therefore, as shown in Fig. \ref{fig:lstm} (a), in order to capture the contextual information at multiple scales, a new CNN module (mConv) with multi-scale filters adding to the deconstructed ConvLSTM, which is used to replace the general convolutional structure of the input modulation gate. This proposed new ConvLSTM is called multi-scale deconstructed ConvLSTM (MSD-ConvLSTM). As an example in Fig. \ref{fig:lstm} (b), in mConv, besides the basic kernel size $K\geq 3$, $K-2$ and $K+2$ are applied as the size of the other additive convolutional kernels. The calculated results of these multi-scale filters are concatenated as the output of the input modulation gate.  The output $\mathbf{g}_{t}$ can be formulated as follows:
\begin{align}
\label{g}
\mathbf{g}_{t}&=\phi([\mathbf{W}_{xc}^{s}\ast\mathbf{X}_{t}, \mathbf{W}_{xc}^{m}\ast\mathbf{X}_{t}, \mathbf{W}_{xc}^{l}\ast\mathbf{X}_{t}] \nonumber\\
&+ [\mathbf{W}_{hc}^{s}\ast\mathbf{H}_{t}, \mathbf{W}_{hc}^{m}\ast\mathbf{H}_{t}, \mathbf{W}_{hc}^{l}\ast\mathbf{H}_{t}] + \mathbf{b}_{c}),
\end{align}
where $\mathbf{W}_{xc}^{s}$, $\mathbf{W}_{xc}^{m}$, $\mathbf{W}_{xc}^{l}$ denote the small, middle and large weight matrixes of the input; $\mathbf{W}_{hc}^{s}$, $\mathbf{W}_{hc}^{m}$, $\mathbf{W}_{hc}^{l}$ denote the small, middle and large weight matrixes of the hidden state; the channel size of the small, middle and large convolutional output is $\frac{1}{4}C_h$, $\frac{2}{4}C_h$ and $\frac{1}{4}C_h$, respectively. The remaining operations of MSD-ConvLSTM are computed below:
\begin{eqnarray}
\label{LSTM4}
\bar{\bm{x}}_{t}\hspace{-0.2cm}&=&\hspace{-0.2cm}GAP(\mathbf{X}_{t}),\nonumber\\
\bar{\bm{h}}_{t-1}\hspace{-0.2cm}&=&\hspace{-0.2cm}GAP(\mathbf{H}_{t-1}),\\
\bm{i}_{t}^{c}\hspace{-0.2cm}&=&\hspace{-0.2cm}\mathbf{W}_{xi}^{c}\bar{\bm{x}}_{t} + \mathbf{W}_{hi}^{c}\bar{\bm{h}}_{t-1} + \mathbf{b}_{i},\nonumber\\
\mathbf{i}_{t}^{s}\hspace{-0.2cm}&=&\hspace{-0.2cm}\mathbf{W}_{xi}^{s}\ast\mathbf{X}_{t-1} + \mathbf{W}_{hi}^{s}\ast\mathbf{H}_{t-1},\nonumber\\
\mathbf{i}_{t}\hspace{-0.2cm}&=&\hspace{-0.2cm}\sigma(\bm{i}_{t}^{c}\circ\mathbf{i}_{t}^{s}),\\
\bm{f}_{t}^{c}\hspace{-0.2cm}&=&\hspace{-0.2cm}\mathbf{W}_{xf}^{c}\bar{\bm{x}}_{t} + \mathbf{W}_{hf}^{c}\bar{\bm{h}}_{t-1} + \mathbf{b}_{f},\nonumber\\
\mathbf{f}_{t}^{s}\hspace{-0.2cm}&=&\hspace{-0.2cm}\mathbf{W}_{xf}^{s}\ast\mathbf{X}_{t-1} + \mathbf{W}_{hf}^{s}\ast\mathbf{H}_{t-1},\nonumber\\
\mathbf{f}_{t}\hspace{-0.2cm}&=&\hspace{-0.2cm}\sigma(\bm{f}_{t}^{c}\circ\mathbf{f}_{t}^{s}),\\
\bm{o}_{t}^{c}\hspace{-0.2cm}&=&\hspace{-0.2cm}\mathbf{W}_{xo}^{c}\bar{\bm{x}}_{t} + \mathbf{W}_{ho}^{c}\bar{\bm{h}}_{t-1} + \mathbf{b}_{o},\nonumber\\
\mathbf{o}_{t}^{s}\hspace{-0.2cm}&=&\hspace{-0.2cm}\mathbf{W}_{xo}^{s}\ast\mathbf{X}_{t-1} + \mathbf{W}_{ho}^{s}\ast\mathbf{H}_{t-1},\nonumber\\
\mathbf{o}_{t}\hspace{-0.2cm}&=&\hspace{-0.2cm}\sigma(\bm{o}_{t}^{c}\circ\mathbf{o}_{t}^{s}),\\
\mathbf{C}_{t}\hspace{-0.2cm}&=&\hspace{-0.2cm}\bm{f}_{t} \circ \mathbf{C}_{t-1} + \bm{i}_{t} \circ \mathbf{g}_{t},\nonumber\\
\mathbf{H}_{t}\hspace{-0.2cm}&=&\hspace{-0.2cm}\bm{o}_{t} \circ \phi(\mathbf{C}_{t}),
\end{eqnarray}

Therefore, the computational cost is far less than the original ConvLSTM. The parameter size of MSD-ConvLSTM is calculated as:
\begin{align}
\label{msd}
\Theta_{MSD-ConvLSTM} &= [(K\times K) \times (C_h + 3)+5\times C_h] \nonumber\\
&\times(C_x+C_h).
\end{align}
where the numbers of parameters of the original ConvLSTM are greatly reduced.

\subsection{The Architecture of Precipitation Forecasting}

Precipitation is a kind of continuous meteorology influenced by many meteorological elements. Although these factors will affect future rainfall in the corresponding area, according to \cite{andersson2007analysis}, the meteorological elements of relative humidity, air temperature, U component of wind and V component of wind are the main determinants of the rainfall. Therefore, in this paper, the four elements are collected as the inputs of our model. On account of the spatiotemporal property of the meteorological data, an encoder-decoder architecture of CNN+ConvLSTM is adopted to process the meteorological elements. The structure can be seen in Fig. \ref{fig:model}.

\subsubsection{The Encoder}

To extract feature of high performance, a deep CNN is adopted as the structure of the encoder. Due to current precipitation condition is gradually determined by the past period of meteorological conditions. Therefore, for a meteorological element, a sequence of its data is inputted in chronological order to the CNN encoder. Specifically, the continuous inputs of relative humidity, air temperature, U component of wind and V component of wind are defined as $R = \{r_1,r_2,\ldots,r_T\}$, $A = \{a_1,a_2,\ldots,a_T\}$, $U = \{u_1,u_2,\ldots,u_T\}$ and $V = \{v_1,v_2,\ldots,V_T\}$, respectively. The $T$ denotes the length of the sequence. At time step $t$, the feature of each meteorological element is extracted by its CNN encoder. The process is:
\begin{eqnarray}
\mathbf{r}_{t}\hspace{-0.2cm}&=&\hspace{-0.2cm}\text{CNN}(r_t), \mathbf{a}_{l}=\text{CNN}(a_t), \nonumber\\
\mathbf{u}_{t}\hspace{-0.2cm}&=&\hspace{-0.2cm}\text{CNN}(u_t), 
\mathbf{v}_{t}=\text{CNN}(v_t).
\end{eqnarray}

Because of the modes between different meteorological elements are different, the CNN parameters of different meteorological elements are not shared.

\subsubsection{The Decoder}

MSD-ConvLSTM is adopted as the decoder. At time step $t$, the extracted meteorological features of $\mathbf{r}_{t}$, $\mathbf{a}_{t}$, $\mathbf{u}_{t}$ and $\mathbf{v}_{t}$ are concatenated as the input $\mathbf{X}_{t}$ of the MSD-ConvLSTM. The calculating of MSD-ConvLSTM is going from 1 to $T$. The total updating procedure is written below:
\begin{eqnarray}
\mathbf{X}_{t} \hspace{-0.2cm}&=&\hspace{-0.2cm} [\mathbf{r}_{t}, \mathbf{a}_{t}, \mathbf{u}_{t}, \mathbf{v}_{t}],\nonumber\\
\mathbf{H}_{t} \hspace{-0.2cm}&=&\hspace{-0.2cm} \text{MSD-ConvLSTM}(\mathbf{X}_{t}), \quad t \in \{1,\ldots, T\},
\end{eqnarray}
where $\mathbf{H}_{t}$, the current output of $\text{MSD-ConvLSTM}$, is related to the precipitation of the next time step. Finally, $\mathbf{H}_{T}$ is applied to forecast the precipitation during $T$ to $T+1$.

\subsection{The Learning of Precipitation Forecasting}

The output at the final time step of the decoder is imported to the pixel-level classifier. The classifier is composed of two $3\times 3$ convolution layers to refine the features and a simple bilinear upsampling layer to uniform the spatial size with the label. The channel size of the final feature is the same with the categories of precipitation forecasting. The output of a softmax function is inputted to the pixel-wise cross-entropy loss to train the precipitation forecasting model. It penalizes pixel-wise predictions independently.  Given predicted categorical probability $p(y^i|x^i)$ at pixel $i$ w.r.t. its ground truth categorical label $y^i$. Therefore, the total loss at pixel $i$ is written as follows:
\begin{align}
\label{loss}
L = -\log(p(y^i|x^i; \Theta)),
\end{align}
where $\Theta$ is the parameters of the proposed model.

\begin{table}[h!]
	\centering
	\vspace{-0.3cm}	
	\caption{Precipitation classification standards and statistics of the EC and CM datasets.}	
	\label{tab:table0}
	\begin{tabular}{ p{2.1cm}P{2.5cm}P{1cm}P{1cm}}
		
		\toprule
		\multirow{2}{*}{Classification} & \multirow{2}{*}{Precipitation(mm)} & \multicolumn{2}{c}{Proportion(\%)}\\
		\cline{3-4}
		&&EC dataset&CM dataset\\
		\midrule
		No Rain &  $(-\infty, 0.01)$ & $50.90$ &  $90.72$  \\
		Light Rain  & $[0.01, 3)$ & $12.14$ & \multirow{4}{*}{$9.28$}  \\
		Moderate Rain  & $[3, 11)$ & $12.34$ &   \\
		Heavy Rain  & $[11, 25)$ & $10.74$ &   \\
		Rainstorm  & $[25, +\infty)$ & $13.88$ &   \\
		\bottomrule
	\end{tabular} 
	\vspace{-0.3cm}	
\end{table}

\section{Experiments}

\subsection{Datasets}

We report the results of our method on two datasets. The two different datasets named EC and CM are collected by European Centre for Medium-Range Weather Forecasts and China Meteorological Administration Public Meteorological Service Centre, respectively. In the datasets, the numerical values of meteorological elements and precipitation in a certain area of time are provided. The classification standards and statistics of the two datasets can be seen in Table \ref{tab:table0}. Because of the data imbalance, the classification of the CM dataset is set to rain or not rain. And the details of these datasets can be seen as follows:

\subsubsection{EC Dataset}
EC dataset\footnote{\url{https://apps.ecmwf.int/datasets/data/interim-full-daily/levtype=pl/}\label{url1}} provides the numerical values of meteorological elements in worldwide every three hours and reports the numerical values of precipitation during every three hours. The sources of meteorological elements in EC dataset include most elements related to rainfall. Each data of meteorological element is a three dimensional high altitude data which is collected in 37 layers with the atmospheric pressure changes. The data of precipitation is a two-dimensional matrix to label the observations in the same region. The value of every point in the dataset is the average physical quantities of a local region at certain observation. The spatial resolution of all the sources is $0.125^\circ \times 0.125^\circ$. In experiment, we choose the region of East Asia (from N55, W70 to S0, E140) in the dataset as the inputs. The size of this region is $441\times 561$. The data from 2011 to 2016 is adopted as the experimental data. Considering the continuity of time, the data from 2012 to 2015 is used for training, the validation data contains the data in 2011 and the test data includes the data in 2016.

\begin{table*}[h!]
	\centering	
	\vspace{-0.3cm}	
	\caption{Comparison among the original ConvLSTM and the variants on the EC dataset. Channel Size denotes the feature of the hidden states. Parameter Size is calculated from the LSTM in these models. The Running Time is the average testing time of the decoder module in testing. }	
	\label{tab:table1}
	\begin{tabular}{ p{3.9cm}P{1.5cm}P{1.5cm}P{2cm}P{3cm}P{3cm}}
		
		\toprule
		Model &Acc&mIoU&Channel Size &Parameter Size&Running Time (ms)\\
		\midrule
		ConvLSTM  &  $0.9007$ & $0.8192$ &  $128$ & $3391488$ &  $7.922$  \\
		FC-ConvLSTM  & $0.8998$ & $0.8178$ & $128$ & $1130496$ & $5.043$  \\
		sConv-ConvLSTM  & $0.9001$ &  $0.8183$ & $128$ & \boldmath $867744$ & \boldmath $4.585$  \\
		Deconstructed-ConvLSTM  &  $0.9007$ & $0.8191$ &  $128$ & $1150368$ &  $5.155$\\
		MSD-ConvLSTM  & \boldmath $0.9026$ & \boldmath $0.8217$ &  $128$ & $1338784$ &  $5.605$ \\
		\bottomrule
	\end{tabular} 
	\vspace{-0.3cm}	
\end{table*}

\begin{table*}[h!]
	\centering	
	\vspace{-0.3cm}	
	\caption{Comparison with the state-of-the-art models on the EC and CM datasets to predict rain or not.}	
	\label{tab:table2}
	\begin{tabular}{ p{5.5cm}P{2cm}P{2cm}P{2cm}P{2cm}}
		
		\toprule
		\multirow{2}{*}{Model} & \multicolumn{2}{c}{EC dataset} & \multicolumn{2}{c}{CM dataset}\\
		\cline{2-5}
		&Acc&mIoU&Acc&mIoU\\
		\midrule
		SVM  &  $0.7125$ & $0.6452$ &  $0.8021$ & $0.5173$   \\
		FCN  & $0.8602$ & $0.7540$ & $0.9042$ & $0.6155$   \\
		DeepLabv3+  & $0.8716$ &  $0.7723$ & $0.9127$ &  $0.6453$   \\
		\midrule
		DeepLabv3+ plus MSD-ConvLSTM  & \boldmath $0.9026$ & \boldmath $0.8217$ & \boldmath $0.9273$ & \boldmath $0.6695$  \\
		\bottomrule
	\end{tabular} 
	\vspace{-0.3cm}	
\end{table*}

\begin{table}[h!]
	\centering	
	\vspace{-0.3cm}	
	\caption{Comparisons on the EC dataset for multiple levels of precipitation forecasting.}	
	\label{tab:table3}
	\begin{tabular}{ p{5.2cm}P{0.9cm}P{0.9cm}}
		
		\toprule
		Model & Acc&mIoU \\
		\midrule
		SVM  &  $0.5326$ & $0.1653$    \\
		FCN  & $0.6530$ & $0.3371$    \\
		DeepLabv3+  & $0.7022$ &  $0.4043$    \\
		\midrule
		DeepLabv3+ plus MSD-ConvLSTM  & \boldmath $0.7439$ & \boldmath $0.4849$  \\
		\bottomrule
	\end{tabular} 
	\vspace{-0.3cm}	
\end{table}

\subsubsection{CM Dataset}
CM dataset only provides the numerical values of precipitation during every six hours. The dimension of the precipitation label is two as well. The coverage area of the label is East Asia (from N55, W70 to S0, E140) and the spatial resolution is the same with EC dataset. Therefore, the numerical values of meteorological elements of the corresponding area in EC dataset are taken as the inputs to forecast the label in CM dataset. This dataset reports the precipitation from 2011 to 2015. In experiment, the data from 2012 to 2014 is used for training, the validation data contains the data in 2011 and the data in 2015 applied to test.

\begin{figure*}[t]
	\begin{center}
		\includegraphics[scale=.65]{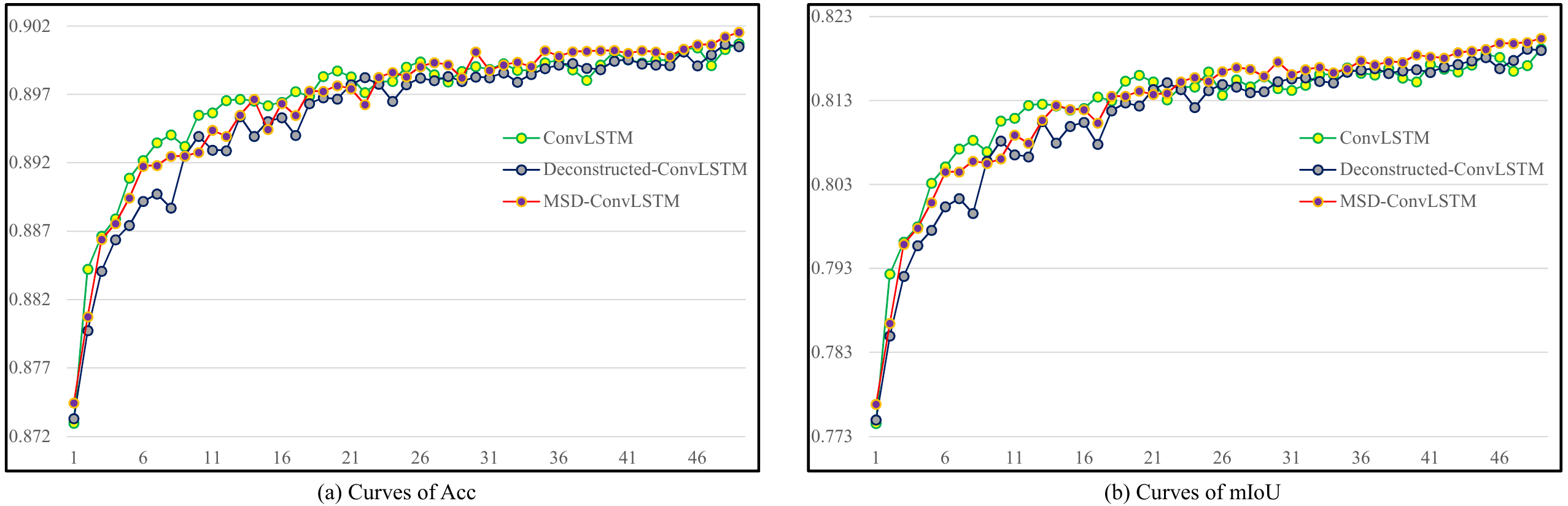}
		
	\end{center}
    \vspace{-0.5cm}	
	\caption{The Acc and mIoU curves in 50 epochs of the ConvLSTM, Deconstructed-ConvLSTM and MSD-ConvLSTM based models on the EC validation set for rain or not prediction. The values are presented on the y-axis.}
	\label{fig:acc}
	\vspace{-0.3cm}	
\end{figure*}

\begin{figure*}[h!]
	\begin{center}
		\includegraphics[scale=.75]{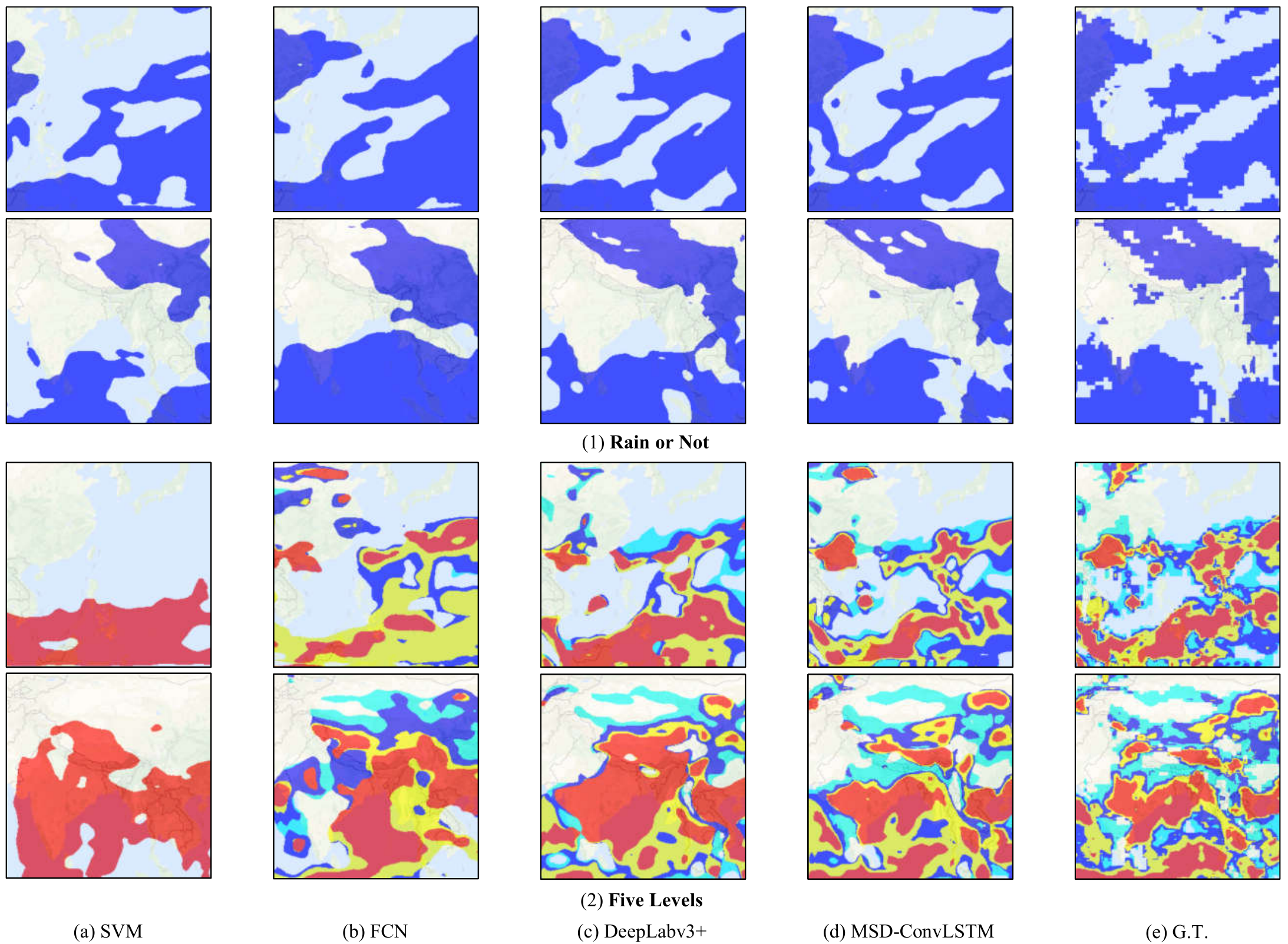}
		
	\end{center}
    \vspace{-0.5cm}	
	\caption{The visualization of SVM, FCN, DeepLabv3+, MSD-ConvLSTM precipitation forecasting results and groundtruth. The top-2 lines indicate two samples of the rain or not results and the blue region denotes the coverage will be rain. The bottom-2 samples present the forecasting of 5 levels. Wathet is light rain, blue denotes moderate rain, yellow means heavy rain, red is rainstorm and the other means no rain, respectively.}
	\label{fig:vv}
	\vspace{-0.5cm}	
\end{figure*}

\subsection{Evaluation}

The widely used criteria of the accuracy (Acc) and the mean intersection over union (mIoU) are adopted as the reported metrics in experiment.

\subsection{Implementation Details}

\subsubsection{The Encoder}
DeepLabv3+ \cite{ChenZPSA18} is one of the baseline model in the semantic information extraction of spatial data. The utilized resnet-101 \cite{HeZRS16} part is pre-trained on the ImageNet \cite{KrizhevskySH12} in our experiments. The numerical values of relative humidity, air temperature, U component of wind and V component of wind are respectively imported into different DeepLabv3+ models without parameter sharing. The spatial size of these meteorological elements is $432\times 544$. And the size of each final extracted CNN feature is $152\times 108\times 136$.

\subsubsection{The Decoder}
MSD-ConvLSTM is adopted as the decoder in the final model. The channel size of the hidden states in MSD-ConvLSTM is set to 128. The basic kernel size $K$ is set to be 3. To ensure both performance and efficiency, the length $T$ of the sequence in model is set to be 4. The interval between the adjacent time inputted meteorological elements is six hour. It means that the forecasting precipitation is to predict the rainfall level in the next six hours after the inputs of the $T$-th time step.

The Adamx optimizer \cite{PhuongP19} is adopted in training. The maximum number of epochs of the training is 100. At each epoch, the validation set is used to evaluate the training model, and the best score model is selected for the final testing. All of our experiments are implemented with Pytorch \cite{paszke2017automatic}.

\subsection{Compared Approaches}
To verify the effectiveness of deep learning, SVM \cite{BennettC00} is trained for comparison. The other baseline models like FCN \cite{LongSD15}, DeepLabv3+, etc. without connecting to the recurrent neural network are introduced to prove the performance of the spatiotemporal modeling of our proposed MSD-ConvLSTM.


\subsection{Explorative study}
Explorative studies for the ConvLSTM and the other variants of ConvLSTM are performed in this section to predict the rain or not. The results on the EC dataset are shown in Table \ref{tab:table1}. The difference between these models is the ConvLSTM in the decoder.

The ConvLSTM-based model is adopted as the baseline model. We first investigate the effectiveness of the replacement from convolution to the FC layer of the three gates (FC-ConvLSTM).  From the results, compared to ConvLSTM, the parameter size of FC-ConvLSTM is reduced $\frac{2}{3}$, but the performance of FC-ConvLSTM is reduced very limited. The running time of FC-ConvLSTM saves $2.879$ ms to ConvLSTM for each iteration. It indicates that the effect of the convolutional structures is not always remarkable.

Then, the effectiveness of the single channel convolution (sConv) is studied. Replacing the convolutional structure by sConv can further descend the parameter size which is about $\frac{1}{4}$ the size of ConvLSTM. Compared to FC-ConvLSTM, sConv-ConvLSTM based model achieves higher performance with less running time costing. The capture of the contextual information between the pixel-level feature can make the gates better selectively measure the inputs.

The effect of the Deconstructed-ConvLSTM is taken into study. From the results of deconstructed-ConvLSTM, the Deconstructed-ConvLSTM based model achieves the same performance with ConvLSTM based model on the Acc and mIoU metrics. Meanwhile, the parameter size of deconstructed-ConvLSTM is about $\frac{1}{3}$ the size of ConvLSTM and the running time reduces $2.767$ ms to ConvLSTM for each iteration. The comparison demonstrate the effectiveness of the deconstructed operation.

Finally, we explore the influence of the multiple scale filters. From the results, the proposed MSD-ConvLSTM shows a better performance than other variants. It demonstrates that multiple scale filters can help to capture more rich contextual information. And comparing to ConvLSTM, MSD-ConvLSTM is more efficient and effective.

\subsection{Quantitative Comparisons}
Table \ref{tab:table2} shows the comparison results of rain or not with other state-of-the-art models on the EC and CM datasets. The performance of the SVM model is the poorest compared with the deep learning models, which indicates deep learning has better modeling capability. Comparing with the pure CNN models like FCN and DeepLabv3+, the introduction of RNNs for spatiotemporal modeling can capture temporal contextual information to improve the performance of precipitation forecasting. From the results, the proposed encoder-decoder of ``DeepLabv3+ plus MSD-ConvLSTM" achieves the highest performance on the two datasets. It should be noticed that the imbalance of the CM dataset makes it hard to capture enough rainfall information, but our proposed model still achieves the highest performance.

Moreover, Table \ref{tab:table3} shows the comparisons of multiple rain levels on the EC dataset. It is seen that the proposed encoder-decoder of ``DeepLabv3+ plus MSD-ConvLSTM" achieves the highest performance among the other architectures.

\subsection{Visualized Analysis}

In Fig. \ref{fig:acc}, the Acc and mIoU curves in 50 epochs of the ConvLSTM, Deconstructed-ConvLSTM and MSD-ConvLSTM based models on the EC validation set are visualized to verify the effectiveness of the deconstruction operation and multi-scale filters. From the curves, in the early times of learning, ConvLSTM based model shows better performance. As the training epoch increased, Deconstructed-ConvLSTM and MSD-ConvLSTM based models show better ability to fuse the contextual information of features. It indicates that the deconstruction operation and multi-scale filters are effective.

The precipitation forecasting results of SVM, FCN, DeepLabv3+, DeepLabv3+ plus MSD-ConvLSTM models and groundtruth are visualized in Fig. \ref{fig:vv}. The results are obtained from the EC dataset. From the examples, it can be easy concluded that the forecasting of our method is much more approaching to the groundtruth.

\section{Conclusion}
In this paper, the multi-scale deconstructed ConvLSTM (MSD-ConvLSTM) is proposed with small quantity of parameters and powerful capacity of fusion. From the evaluation results, the deconstruction of the original convolutional structure and the introduction of the multi-scale filters can improve the efficiency and effectiveness of ConvLSTM. Furthermore, to effectively combine NWP with deep learning, an encoder-decoder architecture with the MSD-ConvLSTM is proposed for precipitation forecasting. The numerical values of multiple meteorological elements are encoded by a deep CNN and decoded by the MSD-ConvLSTM to fuse the spatiotemporal information from different elements at different time steps. The experiments demonstrate that the architecture provides an efficient and effective data-driven way in precipitation forecasting. Moreover, the conditions of precipitation show different levels of connection with different meteorological elements at different times. For future work, we will investigate the function of the attention mechanism in the precipitation forecasting.

{\small
		\bibliographystyle{aaai}
		\bibliography{weather}

\begin{thebibliography}{}

\bibitem[\protect\citeauthoryear{Alahi \bgroup et al\mbox.\egroup
  }{2016}]{AlahiGRRLS16}
Alahi, A.; Goel, K.; Ramanathan, V.; Robicquet, A.; Li, F.; and Savarese, S.
\newblock 2016.
\newblock Social {LSTM:} human trajectory prediction in crowded spaces.
\newblock In {\em {IEEE} Conference on Computer Vision and Pattern Recognition
  (CVPR)},  961--971.

\bibitem[\protect\citeauthoryear{Andersson \bgroup et al\mbox.\egroup
  }{2007}]{andersson2007analysis}
Andersson, E.; H{\'o}lm, E.; Bauer, P.; Beljaars, A.; Kelly, G.; McNally, A.;
  Simmons, A.; Th{\'e}paut, J.-N.; and Tompkins, A.
\newblock 2007.
\newblock Analysis and forecast impact of the main humidity observing systems.
\newblock {\em Quarterly Journal of the Royal Meteorological Society}
  133(627):1473--1485.

\bibitem[\protect\citeauthoryear{Bennett and Campbell}{2000}]{BennettC00}
Bennett, K.~P., and Campbell, C.
\newblock 2000.
\newblock Support vector machines: Hype or hallelujah?
\newblock {\em {SIGKDD} Explorations} 2(2):1--13.

\bibitem[\protect\citeauthoryear{Campbell and
  Diebold}{2005}]{campbell2005weather}
Campbell, S.~D., and Diebold, F.~X.
\newblock 2005.
\newblock Weather forecasting for weather derivatives.
\newblock {\em Journal of the American Statistical Association} 100(469):6--16.

\bibitem[\protect\citeauthoryear{Chen and Lai}{2011}]{chen2011comparison}
Chen, L., and Lai, X.
\newblock 2011.
\newblock Comparison between arima and ann models used in short-term wind speed
  forecasting.
\newblock In {\em {IEEE} Asia-Pacific Power and Energy Engineering Conference},
   1--4.

\bibitem[\protect\citeauthoryear{Chen \bgroup et al\mbox.\egroup
  }{2018}]{ChenZPSA18}
Chen, L.; Zhu, Y.; Papandreou, G.; Schroff, F.; and Adam, H.
\newblock 2018.
\newblock Encoder-decoder with atrous separable convolution for semantic image
  segmentation.
\newblock In {\em European Conference on Computer Vision (ECCV)},  833--851.

\bibitem[\protect\citeauthoryear{Dai \bgroup et al\mbox.\egroup
  }{2016}]{DaiLHS16}
Dai, J.; Li, Y.; He, K.; and Sun, J.
\newblock 2016.
\newblock {R-FCN:} object detection via region-based fully convolutional
  networks.
\newblock In {\em Advances in Neural Information Processing Systems (NeruIPS)},
   379--387.

\bibitem[\protect\citeauthoryear{Gneiting and
  Raftery}{2005}]{gneiting2005weather}
Gneiting, T., and Raftery, A.~E.
\newblock 2005.
\newblock Weather forecasting with ensemble methods.
\newblock {\em Science} 310(5746):248--249.

\bibitem[\protect\citeauthoryear{Gope \bgroup et al\mbox.\egroup
  }{2016}]{gope2016early}
Gope, S.; Sarkar, S.; Mitra, P.; and Ghosh, S.
\newblock 2016.
\newblock Early prediction of extreme rainfall events: a deep learning
  approach.
\newblock In {\em Industrial Conference on Data Mining (ICDM)},  154--167.

\bibitem[\protect\citeauthoryear{Graves, Mohamed, and
  Hinton}{2013}]{GravesMH13}
Graves, A.; Mohamed, A.; and Hinton, G.
\newblock 2013.
\newblock Speech recognition with deep recurrent neural networks.
\newblock In {\em {IEEE} International Conference on Acoustics, Speech and
  Signal Processing (ICASSP)},  6645--6649.

\bibitem[\protect\citeauthoryear{Hall, Brooks, and
  Doswell~III}{1999}]{hall1999}
Hall, T.; Brooks, H.~E.; and Doswell~III, C.~A.
\newblock 1999.
\newblock Precipitation forecasting using a neural network.
\newblock {\em Weather and forecasting} 14(3):338--345.

\bibitem[\protect\citeauthoryear{He \bgroup et al\mbox.\egroup
  }{2016}]{HeZRS16}
He, K.; Zhang, X.; Ren, S.; and Sun, J.
\newblock 2016.
\newblock Deep residual learning for image recognition.
\newblock In {\em {IEEE} Conference on Computer Vision and Pattern Recognition
  (CVPR)},  770--778.

\bibitem[\protect\citeauthoryear{Hern{\'{a}}ndez \bgroup et al\mbox.\egroup
  }{2016}]{HernandezSJCD16}
Hern{\'{a}}ndez, E.; S{\'{a}}nchez{-}Anguix, V.; Juli{\'{a}}n, V.;
  C{\'{a}}mara, J.~P.; and Duque, N.
\newblock 2016.
\newblock Rainfall prediction: {A} deep learning approach.
\newblock In {\em Hybrid Artificial Intelligent Systems (HAIS)},  151--162.

\bibitem[\protect\citeauthoryear{Jain \bgroup et al\mbox.\egroup
  }{2016}]{JainZSS16}
Jain, A.; Zamir, A.; Savarese, S.; and Saxena, A.
\newblock 2016.
\newblock Structural-rnn: Deep learning on spatio-temporal graphs.
\newblock In {\em {IEEE} Conference on Computer Vision and Pattern Recognition
  (CVPR)},  5308--5317.

\bibitem[\protect\citeauthoryear{Krizhevsky, Sutskever, and
  Hinton}{2012}]{KrizhevskySH12}
Krizhevsky, A.; Sutskever, I.; and Hinton, G.
\newblock 2012.
\newblock Imagenet classification with deep convolutional neural networks.
\newblock In {\em Conference on Neural Information Processing Systems
  (NeruIPS)},  1106--1114.

\bibitem[\protect\citeauthoryear{Long, Shelhamer, and Darrell}{2015}]{LongSD15}
Long, J.; Shelhamer, E.; and Darrell, T.
\newblock 2015.
\newblock Fully convolutional networks for semantic segmentation.
\newblock In {\em {IEEE} Conference on Computer Vision and Pattern Recognition
  (CVPR)},  3431--3440.

\bibitem[\protect\citeauthoryear{Maqsood, Khan, and
  Abraham}{2004}]{MaqsoodKA04}
Maqsood, I.; Khan, M.; and Abraham, A.
\newblock 2004.
\newblock An ensemble of neural networks for weather forecasting.
\newblock {\em Neural Computing and Applications} 13(2):112--122.

\bibitem[\protect\citeauthoryear{Navon}{2009}]{navon2009data}
Navon, I.~M.
\newblock 2009.
\newblock Data assimilation for numerical weather prediction: a review.
\newblock In {\em Data assimilation for atmospheric, oceanic and hydrologic
  applications}.
\newblock  21--65.

\bibitem[\protect\citeauthoryear{Paszke \bgroup et al\mbox.\egroup
  }{2017}]{paszke2017automatic}
Paszke, A.; Gross, S.; Chintala, S.; Chanan, G.; Yang, E.; DeVito, Z.; Lin, Z.;
  Desmaison, A.; Antiga, L.; and Lerer, A.
\newblock 2017.
\newblock Automatic differentiation in pytorch.
\newblock In {\em NeruIPS Workshop}.

\bibitem[\protect\citeauthoryear{Phuong and Phong}{2019}]{PhuongP19}
Phuong, T.~T., and Phong, L.~T.
\newblock 2019.
\newblock On the convergence proof of amsgrad and a new version.
\newblock {\em {IEEE} Access} 7:61706--61716.

\bibitem[\protect\citeauthoryear{Richardson}{2007}]{richardson2007weather}
Richardson, L.~F.
\newblock 2007.
\newblock {\em Weather prediction by numerical process}.
\newblock Cambridge University Press.

\bibitem[\protect\citeauthoryear{Sapankevych and Sankar}{2009}]{SapankevychS09}
Sapankevych, N., and Sankar, R.
\newblock 2009.
\newblock Time series prediction using support vector machines: {A} survey.
\newblock {\em {IEEE} Comp. Int. Mag.} 4(2):24--38.

\bibitem[\protect\citeauthoryear{Shi \bgroup et al\mbox.\egroup
  }{2015}]{ShiCWYWW15}
Shi, X.; Chen, Z.; Wang, H.; Yeung, D.; Wong, W.; and Woo, W.
\newblock 2015.
\newblock Convolutional {LSTM} network: {A} machine learning approach for
  precipitation nowcasting.
\newblock In {\em Advances in Neural Information Processing Systems (NeruIPS)},
   802--810.

\bibitem[\protect\citeauthoryear{Sloughter \bgroup et al\mbox.\egroup
  }{2007}]{sloughter2007}
Sloughter, J. M.~L.; Raftery, A.~E.; Gneiting, T.; and Fraley, C.
\newblock 2007.
\newblock Probabilistic quantitative precipitation forecasting using bayesian
  model averaging.
\newblock {\em Monthly Weather Review} 135(9):3209--3220.

\bibitem[\protect\citeauthoryear{Tolstykh and Frolov}{2005}]{tolstykh2005some}
Tolstykh, M., and Frolov, A.
\newblock 2005.
\newblock Some current problems in numerical weather prediction.
\newblock {\em Izvestiya Atmospheric and Oceanic Physics} 41(3):285--295.

\bibitem[\protect\citeauthoryear{Zhang \bgroup et al\mbox.\egroup
  }{2018a}]{0005DSZWTA18}
Zhang, H.; Dana, K.~J.; Shi, J.; Zhang, Z.; Wang, X.; Tyagi, A.; and Agrawal,
  A.
\newblock 2018a.
\newblock Context encoding for semantic segmentation.
\newblock In {\em {IEEE} Conference on Computer Vision and Pattern Recognition
  (CVPR)},  7151--7160.

\bibitem[\protect\citeauthoryear{Zhang \bgroup et al\mbox.\egroup
  }{2018b}]{0010ZMSSB18}
Zhang, L.; Zhu, G.; Mei, L.; Shen, P.; Shah, S. A.~A.; and Bennamoun, M.
\newblock 2018b.
\newblock Attention in convolutional {LSTM} for gesture recognition.
\newblock In {\em Advances in Neural Information Processing Systems (NeruIPS)},
   1957--1966.

\bibitem[\protect\citeauthoryear{Zheng \bgroup et al\mbox.\egroup
  }{2014}]{ZhengCWY14a}
Zheng, Y.; Capra, L.; Wolfson, O.; and Yang, H.
\newblock 2014.
\newblock Urban computing: Concepts, methodologies, and applications.
\newblock {\em {ACM} {TIST}} 5(3):38:1--38:55.

\end{thebibliography}
}
	
\end{document}